\documentclass[useAMS,usenatbib]{mn2e}
\usepackage{graphicx,amsmath}
\usepackage{amssymb}
\usepackage{color}
\newcommand{\HIa}{H\,{\sevensize{\textbf{I}}}\,\,}
\newcommand{\HeIIa}{He\,{\sevensize{\textbf{II}}}\,\,}
\newcommand{\HeIIb}{He\,\sevensize{\textbf{II}}\large\,\,}
\newcommand{\HeIIIa}{He\,\sevensize{\textbf{III}}\normalsize\,\,}

\title[Effect of UV escape fraction on He~{\sc ii}]{He~{\sc ii} optical depth
and UV escape fraction of galaxies}
\author[Khaire and Srianand]{Vikram Khaire\thanks{E-mail:
vikramk@iucaa.ernet.in} and Raghunathan Srianand \\
IUCAA, Post Bag 4, Pune, India - 411007}
\begin{document}

\date{Accepted for publication in MNRAS Letters on 2013 January 17 }

\pagerange{\pageref{firstpage}--\pageref{lastpage}} \pubyear{2012}

\maketitle

\label{firstpage}

\begin{abstract}
We study the effect of \HIa ionizing photons escaping from 
high-redshift (high-$z$) galaxies have on the \HeIIb ionizing ultraviolet 
background (UVB) radiation. While these photons do not 
directly interact with \HeIIb ions, we show that they play an important role, 
through radiative transport, in modifying the shape of
\HeIIb ionizing part of UVB spectrum. Within the observed range
of UV escape from galaxies, we show that the rapid increase in 
\HeIIb Ly$\alpha$ effective optical depth at $z\sim2.7$ can 
naturally be explained by radiative transport effects. 
Therefore, the relationship between 
a well measured \HeIIb Ly $\alpha$ effective optical depth and the redshift
in the post-\HeIIb reionization era can be used to place 
additional constraints on the redshift evolution of UV 
escape from high-$z$ galaxies. Our study also suggests that the escape fraction of \HIa ionizing photons from galaxies has an important role in  
the fluctuations of the \HeIIb ionizing UVB.
\end{abstract}

\begin{keywords}
Intergalactic medium--cosmology: theory--diffuse radiation.
\end{keywords}

\section{Introduction}

The spectroscopy of $z\ge6$ quasi-stellar objects
\citep[QSOs;][]{Fan06} and cosmic 
microwave background (CMB) polarization observations \citep[][]{Larson11} 
suggest that H~{\sc i} in the intergalactic medium (IGM) became reionized 
at $6\le z\le12$. The reionized IGM at $z\le6$ is believed to be in ionization
equilibrium with the ultraviolet background (UVB) emanating from QSOs
and galaxies \citep[][HM12 hereafter]{HM2012}. The measured temperature of 
the IGM at $z\le3$ \citep[][]{Becker11t}, the claimed excess temperature around 
$z\sim6$ QSOs \citep[][]{Bolton12}, the lack of substantial 
contribution to $\rm E>54.4$ eV 
photons from galaxies and the redshift distribution of QSOs 
\citep[][]{Hopkins} all favor He~{\sc ii} reionization occurs in the range 
$2.5\le z\le 6.0$.

Direct measurements of He~{\sc ii} Lyman $\alpha$ (Ly$\alpha$) 
absorption from the IGM is possible
towards few UV bright high-$z$ QSOs using the \emph{Hubble Space Telescope}
\citep[for summary of observations, see][]{Shull10}.
The cosmic-variance limited available data suggest a rapid evolution 
of the He~{\sc ii} Ly-$\alpha$ effective optical depth ($\tau_{\alpha,{\rm HeII}}$) 
and a large fluctuation in the column density ratio of He~{\sc ii} 
and H~{\sc i} (called  $\eta$) in the range $2.7\le z\le 3.0$. 
The rapid evolution of the $\tau_{\alpha,{\rm HeII}}$
is attributed to the completion of He~{\sc ii} reionization at this epoch
\citep{Furlanetto08,Mcquinn09}. The large fluctuation in $\eta$ over
small scales can be attributed to the following: (i) the small number of bright
QSOs within a typical mean free path ($\lambda_{\rm mfp}$) of
He~{\sc ii} ionizing photons \citep{Fardal,Furlanetto09};
(ii) the large scatter in the QSO spectral index \citep{Shull04};
(iii) the presence of collisionally ionized gas \citep{Muzahid10};
(iv) the small scale radiative transport effects \citep{Maselli05}.
Most theoretical calculations of the He~{\sc ii} optical depth have 
considered only the QSO emissivity and radiative transport, assuming that 
the IGM gas in photoionization equilibrium. While galaxies do not
contribute directly to the He~{\sc ii} ionizing radiation, they
can influence the ionization state of the IGM gas, thereby
affecting the He~{\sc ii} optical depth. 
Here, we explore this issue, using a cosmological radiative transfer 
code similar to HM12 keeping the galaxy contribution as a free
parameter within the range allowed by the observations. 
We show that $\lambda_{\rm mfp}$ is very sensitive
to the escape fraction ($f_{\rm esc}$) of H~{\sc i} 
ionizing photons from galaxies. In Section 2,
we provide details of our calculations 
assuming a ($\Omega_{m}$,$\Omega_{\Lambda}$,$h$)=(0.3, 0.7, 0.7) cosmology.
\section{UVB Calculation}\label{sec.rad_t}

We have calculated the UVB spectrum contributed by QSOs and
galaxies using the standard assumption that each volume element is 
an isotropic emitter and sink \citep[e.g.][]{HM96,Fardal,FG09}. 
The angle and space averaged specific intensity $J_{\nu_{0}}$ 
(in units of 
erg cm$^{\text{-2}}$ s$^{\text{-1}}$ Hz$^{\text{-1}}$ sr$^{\text{-1}}$) 
of diffuse UVB, as seen by an observer at a 
redshift $z_{0}$ and frequency $\nu_{0}$ is given by \citep{HM96}:
%
\begin{equation}\label{rad_t}
J_{\nu_{0}}(z_{0})=\frac{1}{4\pi}\int_{z_{0}}^{\infty}dz\,\frac{dl}{dz}\,\frac{(1+z_{0})^{3}}{(1+z)^{3}}\,\epsilon_{\nu}(z)\, e^{-\tau_{\rm eff}(\nu_{0},\, z_{0},\, z)}\,.
\end{equation}
%
Here, $\nu=\nu_{0}(1+z)/(1+z_{0})$ is the frequency of emitted radiation at 
redshift $z$, $\epsilon_{\nu}(z)$ is the proper space-averaged specific 
volume emissivity of radiating sources (QSOs and galaxies),
$\frac{dl}{dz}$ is the proper line element in the 
Friedmann-Robertson-Walker cosmology
and  $\tau_{\rm eff}$ is 
the effective optical depth, which quantifies the attenuation of photons
observed at a frequency $\nu_{0}$  while travelling through the IGM in between
$z$ and $z_0$. 

If we assume that the IGM clouds with neutral hydrogen column density, 
$N_{\rm HI}$, are Poisson-distributed along the line of sight, we can write 
$\tau_{\rm eff}$ as \citep{Paresce},
%
\begin{equation}\label{tau_eff}
\tau_{\rm eff}(\nu_{0}, z_{0}, z)=\int_{z_{0}}^{z}dz'\int_{0}^{\infty}dN_{\rm HI}\frac{\partial^{2}N}{\partial N_{\rm HI}\,\partial z'}[1-e^{-\tau(\nu')}]\,.
\end{equation}
%
Here, $f(N_{\rm HI}, z)=\partial^{2}N/\partial N_{\rm HI}\partial z$, 
is the number of absorbers with $N_{\rm HI}$ per unit redshift 
and column density interval measured at $z$. This is directly
measured with QSO spectroscopy \citep[see][]{Petitjean93}.
Assuming that absorbing clouds are made up of a pure H and He
gas, the continuum optical depth through an individual cloud can be written 
as
%
\begin{equation}\label{tau}
\tau(\nu')=N_{\rm HI}{\sigma_{\rm HI}(\nu')}+N_{\rm HeI}{\sigma_{\rm HeI}(\nu')}+N_{\rm HeII}{\sigma_{\rm HeII}(\nu')}\,\, .
\end{equation}
%
Here, $\nu'=\nu_{0}(1+z')/(1+z_{0})$ and $N_x$ and $\sigma_{x}$ are 
the column density and photoionization cross-section for a species $x$,
respectively. From QSO spectroscopy, we know that $N_{\rm HeI}$
is negligible when $N_{\rm HI}\le10^{17.2}$ cm$^{-2}$. 
Even for Lyman limit systems ($N_{\rm HI} > 10^{17.2}$ cm$^{-2}$), the 
ratio $N_{\rm HeI}/N_{\rm HI}$ is small (see HM12) enough to neglect its 
contribution over the redshift range of interest in our study.
Therefore, Eq.~(\ref{tau}) becomes
\begin{equation}\label{tau_mod}
\tau(\nu)\approx N_{\rm HI}[\sigma_{\rm HI}(\nu)+\eta \, \sigma_{\rm HeII}(\nu)].
\end{equation}
%
In the absence of direct measurements of $f(N_{\rm HeII},z)$, a
knowledge of $\eta$  as a function
of $N_{\rm HI}$ together with  $f(N_{\rm HI}, z)$  allows us to 
calculate the contribution of \HeIIb to the continuum
optical depth.
Under photoionization equilibrium, $N_{\rm HI}$ and $N_{\rm HeII}$
are related through the following quadratic equation \citep{Fardal,FG09,HM2012}:
%
\begin{equation}\label{eta}
\frac{n_{\rm He}}{4n_{\rm H}}\frac{I_{\rm HI}~\tau_{\rm 912, HI}}{(1+{\rm A}~\tau_{\rm 912,HI})}=\tau_{\rm 228, HeII}+
\frac{I_{\rm HeII}~\tau_{\rm 228, HeII}}{(1+{\rm B}~\tau_{\rm 228, HeII})} .
\end{equation}
%
Here, $\tau_{\lambda, x}$ is $N_x\sigma_x(\lambda)$ for the species $x$.
The values of A and B depend on the assumed relationship between
the total hydrogen column density ($N_{\rm H}$) and electron density 
($n_e$). This relationship is obtained for a constant density slab
of gas with a thickness of Jeans length under optically thin
photoionization equilibrium \citep{Schaye01}. Numerical simulations
suggest that such a relationship is valid for log~$N_{\rm H}$$\le$18
\citep[see][]{Rahmati12}. We have considered a series of photoionization
models of plane parallel slab having $N_{\rm H}$-$n_e$ relationship of
\citet{Schaye01} illuminated by a power law source 
using the photoionization code {\sc Cloudy} 
\citep[see][]{Ferland98}. 
We confirm that
the values A = 0.02 and B = 0.25 (as used by HM12)
provide a good fit to the model predictions, and adopt these
values in our calculations.
The quantity $I_{x}$ for the $x^{\rm th}$ species is defined as
$I_{x}=\Gamma_{x}/{n_{e}\alpha_{x}(\rm T)}$, where
$\alpha_{x}(\rm T)$ is the case-A recombination coefficient 
and $\Gamma_{x}$ is photoionization rate of $x^{\rm th}$ species as given by,
%
\begin{equation}\label{Gama}
\Gamma_{x}=\int_{\nu_{x}}^{\infty}d\nu\,4\pi\,\frac{J_{\nu}}{h\nu}\,\sigma_{x}(\nu)\,\,.
\end{equation}
%
Here, $\nu_{x}$ is the ionization threshold frequency for the species $x$.
In all our calculations, we use $\rm T=2\times 10^{4}$ K and the form 
of $n_{e}$ and $f(N_{\rm HI},z)$ as given by HM12.
%
\begin{figure}
	\centering
	\includegraphics[bb=93 362 510 710,width=8.0cm,keepaspectratio,clip=true] {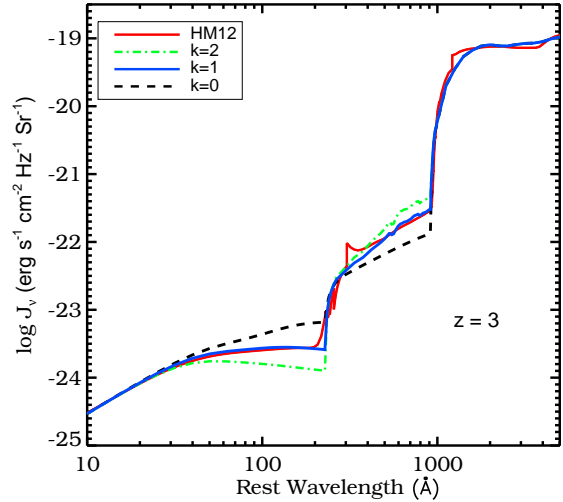}
	\caption{UV background spectrum at $z=3$ from our models
for different values of $k$. The spectrum plotted in red is from HM12.}
\label{uvb}
\end{figure}
%
\subsection{Quasar and galaxy emissivity}\label{sec.emis}
The specific volume emissivity of radiating sources, $\epsilon_{\nu}(z)$, 
is a sum of the emissivity from galaxies, 
$\epsilon_{\nu , \rm G}(z)$, and quasars, $\epsilon_{\nu , \rm Q}(z)$.
We have considered the parametric form to the observed quasar 
co-moving emissivity at 912\AA~, as given in HM12,
%
\begin{equation}\label{Eqso}
\frac{\epsilon_{912, \rm Q}(z)}{(1+z)^{3}}=10^{24.6}\,(1+z)^{4.68}\,\frac{exp(-0.28z)}{exp(1.77z) + 26.3}
\end{equation}
%
in units of ergs Mpc$^{\text{-3}}$ s$^{\text{-1}}$ Hz$^{\text{-1}}$, 
together with the broken power law spectral 
energy distribution (SED), $L_\nu\propto \nu^{-0.44}$ for $\lambda>$1300\AA~and 
$L_\nu\propto \nu^{-1.57}$ for $\lambda<$1300\AA~\citep{vanden,tefler}.
To calculate the co-moving emissivity from galaxies we have taken the 
parametric form of star formation rate density 
(SFRD) used by HM12:
%
\begin{equation}\label{SFRD}
{\rm SFRD}(z)=\frac{6.9\times10^{-3}+0.14(z/2.2)^{1.5}}{1+(z/2.7)^{4.1}}~ {\rm M_{\odot} yr^{-1} Mpc^{-3}}.
\end{equation}
%
The co-moving emissivity of galaxies 
(in units ergs Mpc$^{\text{-3}}$ s$^{\text{-1}}$ Hz$^{\text{-1}}$) is taken to be,
%
\begin{equation}\label{Egal}
\frac{\epsilon_{\nu,\rm G}(z)}{(1+z)^{3}}=C(z)\times {\rm SFRD}(z)\times {l_{\nu}(z, \rm Z)}\,\, .
\end{equation}
%
%
Here, $l_{\nu}(z,\rm Z)$ is  
the specific luminosity of a galaxy produced for every solar mass of
gas having metallicity $\rm Z$ being converted to stars.
We obtain $l_{\nu}(z,{\rm Z})$ using the stellar population synthesis 
model `{\sc Starburst99 v6.0.3}' \citep{Leitherer99} for Z=0.001 and 
Salpeter initial mass function with stellar mass
range 0.1 to 100 M$_{\odot}$.  
SED fitting studies, semi-analytic modeling of luminosity
function and spatial clustering are consistent with
star formation in a typical galaxy lasting for a few 100 Myr \citep[see discussions in]
[and references there in]{Jose12}. Therefore, for simplicity, 
we assume that the star 
formation has lasted for more than 100 Myr. This assumption allows
us to get a linear relationship between the star formation rate and
luminosity. We confirm that $\epsilon_{912,G}$, obtained using this
assumption, is consistent with the 
one using convolution integral (equation 55 of HM12). 
We do not include recombination emissivity and resonant absorption effects in our calculations.

The factor $C(z)$ is used to modify the SED in order 
to take care of the reddening and UV escape. 
For $\lambda >$912\AA, $C(z)$ is $exp(-\tau_\nu)$ with
$\tau_\nu$ being the frequency-dependent dust optical depth, calculated
using extinction law of  \citet{Calzetti} with $\rm R_{V} =3.1$.
The dust correction might depend on the luminosity and $z$ of
individual galaxies \citep[see for example][]{Bouwens12}. 
However, for simplicity, we use a single value at all values of $z$. 
The dust optical depth is chosen to 
have a reddening correction factor of 3 at 1500\AA.

\begin{figure}
	\centering
	\includegraphics[bb=93 362 510 710,width=8.0cm,keepaspectratio,clip=true] {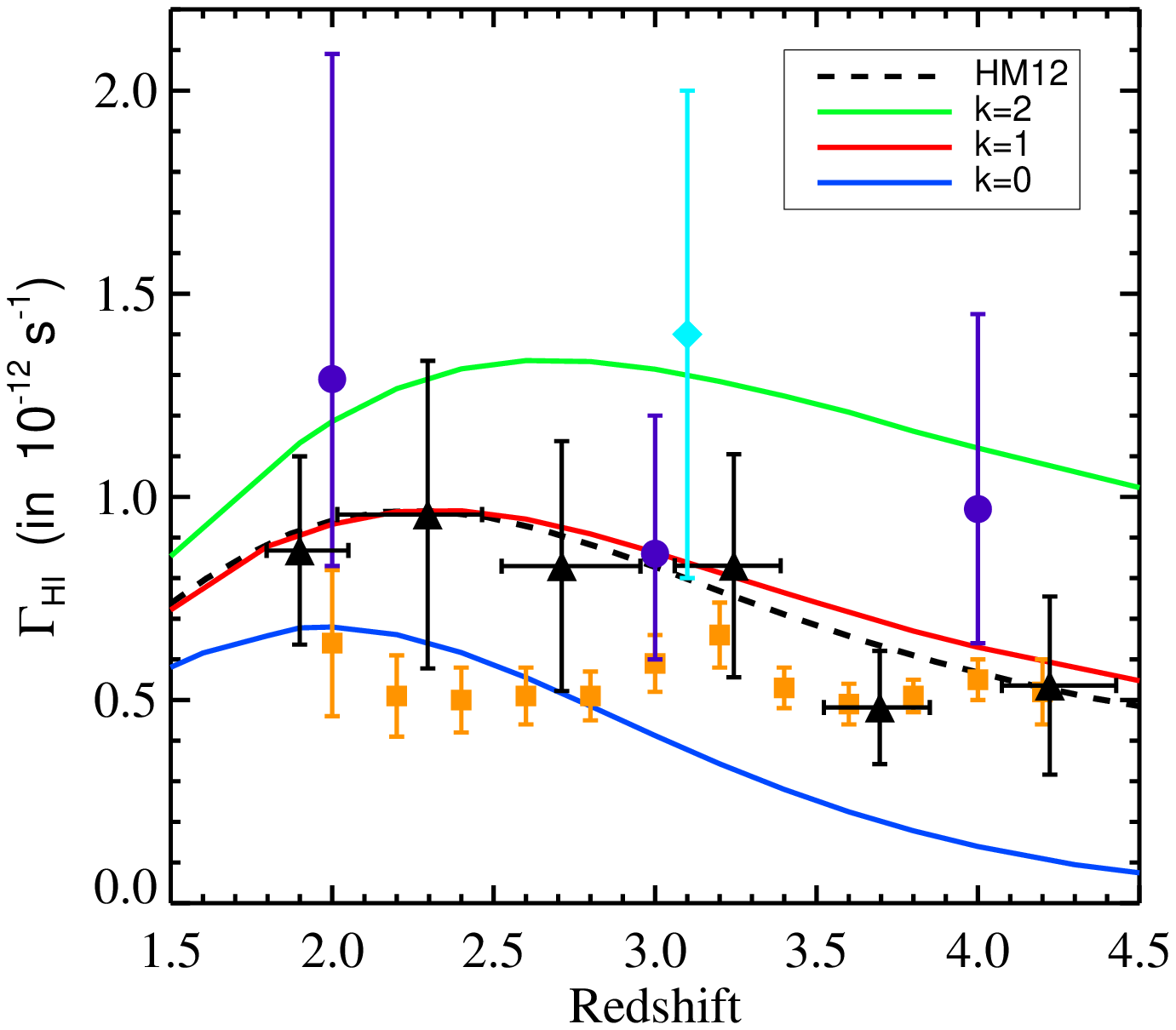}
	\caption{$\Gamma_{\rm HI}$ vs $z$ for k=0 to 2. \emph{Squares}: \citet{FG08}.
\emph{Triangles}: \citet{Becker07} (lognormal model). 
\emph{Circles}: \citet{Bolton07}. \emph{Dimond}: \citet{Nestor12}.}\label{gama_k}
\end{figure}
For wavelength range 228\AA $< \lambda < $912\AA~ we
take $C(z)=f_{\rm esc}$. To take into account the redshift evolution
in the $f_{\rm esc}$ found by \citet{Inoue06}, we adopt the form of HM12 and
use 
\begin{equation}\label{fesc}
C(z)=f_{\rm esc}=k\,\,[3.4\times10^{-4}(1+z)^{3.4}]\,\,\,.
\end{equation}
%
In this case, $C(z)$ corresponds to the absolute escape fraction, 
$f_{\rm esc}$, defined as the ratio of  
escaping Lyman continuum (LyC) flux from a galaxy to the one 
which is intrinsically produced by the stars in it \citep{Leitherer95},
and $k$ is a free parameter that allows
us to change the values of $f_{\rm esc}$. 
For our fiducial star formation model, at $k\,$=$\,1$, 
we obtain UV emissivity from galaxies similar to that of HM12.
The model with  $k\,$=$\,0$ 
corresponds to spectrum contributed by QSOs alone at 
$\lambda <$912\AA.
Note that no reddening correction is applied for 
$\lambda <$ 912\AA~and we simply scale the
unattenuated spectrum by $f_{\rm esc}$. In other words,
we assume that the ionizing photons that are escaping through 
holes in the galaxy. 
We take $C(z)=0$ for $\lambda <$228\AA~because 
galaxies at the redshifts in which we are interested 
do not produce sufficient \HeIIa 
ionizing photons because massive Population-{\sc iii} stars are rare. 
\section{Results and Discussion}
In Fig. \ref{uvb}, we present the UVB spectrum at $z\,$=$\,3$ computed 
using our numerical code for three different values of $k$. 
At $z\,$=$\,3$, $k\,$=$\,1$ and 2 corresponds to  $f_{\rm esc}$ of $\sim$4\%
and $\sim$8\%, respectively. This is well within the inferred range for 
high-$z$ galaxies and 
Ly-$\alpha$ emitters at $z\,$$\sim \,$3
\citep[][]{Shapley06,Iwata09,Boutsia11,Nestor12}.
The range of  $\Gamma_{\rm -12, HI}=0.4-1.3$ predicted  by our models
is also consistent with the range allowed by the
observations (see Fig.~\ref{gama_k}).
For comparison, in Fig. \ref{uvb}, we also plot the 
UVB spectrum generated by HM12. 
It is clear that our model with $k\,$=$\,1$ reproduces
HM12 spectrum very well. The differences in $\Gamma_{\rm HI}$ and
$\Gamma_{\rm HeII}$ between the two codes is much less than the
spread in these values because of the allowed range in $f_{\rm esc}$.
It is clear from Fig.~\ref{uvb} that increasing (decreasing) $k$
increases (decreases) $\Gamma_{\rm HI}$ and decreases (increases)
$\Gamma_{\rm HeII}$. Because $\lambda \le $228\AA~emissivity is not affected by
galaxies, the variation of $\Gamma_{\rm HeII}$ with 
$f_{\rm esc}$ can be attributed to the effect of $f_{\rm esc}$ on 
\HeIIb opacity. Here,
we explore this in more detail.

\subsection{Escape fraction and $\mathbf{\eta}$}\label{sec.fesc}
\begin{figure}
	\centering
	\includegraphics[bb=93 362 510 710,width=8.0cm,keepaspectratio] {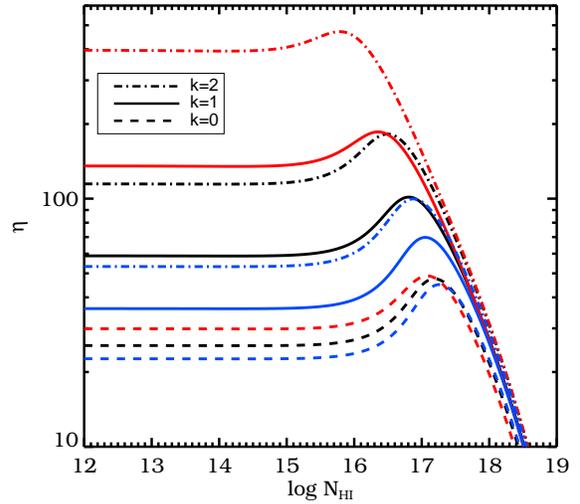}
	\caption{ \small $\eta$ as a function of N$_{\rm HI}$ for redshifts 2 (\emph{blue}), 2.5 (\emph{black}) 
	and 3 (\emph{red}). The dash, solid and dot-dash curves are for $k$ = 0, 1 and 2 respectively.  
}\label{etaz}
\end{figure}
In Fig. \ref{etaz}, we plot $\eta$ versus 
N$_{\rm HI}$ at redshift 2, 2.5 and 3 for different values of $k$.
Two trends are clearly evident for log$(N_{\rm HI})\le 17$: 
(i) for any given $f_{\rm esc}$, $\eta$ increases with increasing $z$
and (ii) at any $z$, $\eta$ increases with increasing $f_{\rm esc}$.
The first trend can be
understood in a simple way for the optically thin 
limit where $\eta 
\propto \Gamma_{\rm HI}/\Gamma_{\rm HeII}$.
Both QSOs and galaxies contribute to $\Gamma_{\rm HI}$ 
but only QSOs contribute to $\Gamma_{\rm HeII}$. Therefore, 
$\eta$ depends on how the 
population of QSOs and galaxies evolve with redshift. For $z\ge2.5$, 
the population 
of QSOs declines rapidly \citep{Ross12}, while that of galaxies remains almost 
the same \citep{Bouwens11}, 
which helps $\eta$ to increase. However, at any $z$, 
by increasing $f_{\rm esc}$,
we are increasing the galaxy contribution to 
$\Gamma_{\rm HI}$, which will further
increase $\eta$. This explains the second trend that we notice in 
Fig. \ref{etaz}.
\emph{We can draw a simple physical picture: a given N$_{\rm HI}$ 
is produced by integrating over a larger column length when $\Gamma_{\rm HI}$
is high, and therefore we obtain 
more N$_{\rm HeII}$}. However, the effects which we
discuss here are obtained by keeping $\Gamma_{\rm HI}$ well withing the range 
allowed by the existing observations (see Fig.~\ref{gama_k}).
\begin{figure}
	\centering
	\includegraphics[bb=93 362 510 710,width=8.0cm,keepaspectratio] {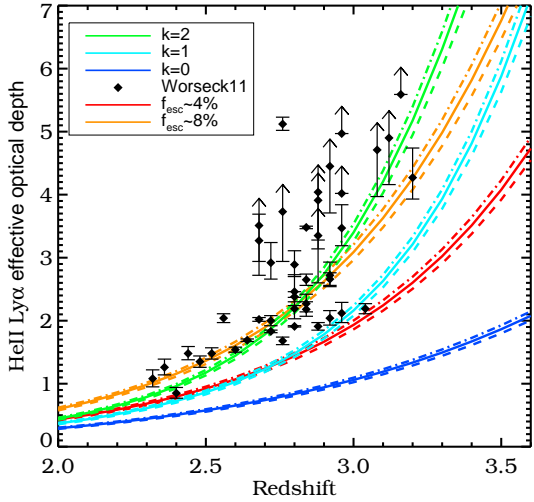}
	\caption{ \small Ly$\alpha$ effective optical depth for He~{\sc ii} as a 
	function of redshift for 
	different f$_{\rm esc}$ and $b$ parameters ($b=28$ \emph{dash curve},
	$b=30$ \emph {solid curve}, $b=32$ \emph{dot dash curve} in km/s). 
	Black diamonds are observations of
	\citet{Worseck11}.}\label{tauHe}
\end{figure}\

\subsection{Ly~$\alpha$ effective optical depth for \HIa and \HeIIa}\label{sec.tau}
Now, we estimate the Ly~$\alpha$ effective optical depth of 
\HIa (i.e., $\tau_{\rm \alpha,HI}$) and \HeIIa (i.e., $\tau_{\rm \alpha,HeII}$)
for a range of $f_{\rm esc}$, and we compare the results 
with the available observations.
The $\tau_{\rm \alpha,HI}$ and $\tau_{\rm \alpha,HeII}$ are given by 
\citep{Paresce,Madau94},
%
\begin{equation}\label{eff_op}
\tau_{\alpha,x}(z)=\frac{1+z}{\lambda_{\alpha, x}}\int_{0}^{\infty}dN_{\rm HI}\,f(N_{\rm HI},z)\, W_{\rm n}\,\,,
\end{equation}
%
where, $\lambda_{\alpha,x}$ is 1215.67\AA~for \HIa, 303.78\AA~for \HeIIa.
Here, $W_{\rm n}$ is the equivalent width of corresponding line,
expressed in wavelength units, given by
%
\begin{equation}\label{wn}
W_{\rm n}=\int_{0}^{\infty}d\lambda\,(1-e^{-y\phi(\lambda)})\,\,\,,
\end{equation}
%
where $y=N_{\rm HI}$ for \HIa, $y=\eta \, N_{\rm HI}$ for \HeIIa and 
$\phi(\lambda)$ is Voigt profile function. 
We find that the observed relationship between $\tau_{\rm \alpha,HI}$ and $z$ 
\citep{Becker12}
is well reproduced if we use $b$ parameter values $b=30\pm2$ kms$^{-1}$.

In Fig. \ref{tauHe}, we plot $\tau_{\rm \alpha,HeII}$ 
versus $z$ for different values of $f_{\rm esc}$, along with observations 
of \citet{Worseck11}. Here, we assume that the non-thermal motions dominate 
the \HeIIb line broadening and used the best-fitting values of $b$ obtained 
for H~{\sc i}.
In all cases $\tau_{\rm \alpha,HeII}$ increases with increasing $z$.
However the rate of increase depends on $f_{\rm esc}$.
Interestingly, the sharp raising trend of 
$\tau_{\rm \alpha,HeII}$ at $z\sim \,$3, seen for $k\,$=$\,1$, 
almost provides a lower envelop to the observations. 
For $k\,$=$\,2$, the model
prediction almost passes through most of the observed mean 
points at $z\le \,$2.7. This means that $f_{\rm esc}$ can not
be much higher than 8\% in this redshift range.
To check whether the rapid rise in $\tau_{\rm \alpha,HeII}$ is a
consequence of the $(1+z)^{3.4}$ evolution assumed for $f_{\rm esc}$ 
(Eq. \ref{fesc}), in Fig.\ref{tauHe} we also plot the
$\tau_{\rm \alpha,HeII}$ computed for redshift-independent $f_{\rm esc}$ 
(i.e. 4\% and 8\%). 
Even in these cases, the $z$ evolution of $\tau_{\rm \alpha,HeII}$ is  
steeper than for $k\,$=$\,0$.
This observed rise and scatter in $\tau_{\rm \alpha,HeII}$ is usually
attributed to the pre-overlap era of \HeIIIa bubbles, assuming that the
\HeIIa re-ionization completes around $z\,$=$\,2.7$~\citep{Dixon09,Shull10,Worseck11}. 
\emph{Our results suggest that the 
observed trend of $\tau_{\rm \alpha,HeII}$ with $z$ can also be produced 
naturally 
by the radiative transfer effects associated with changing $f_{\rm esc}$}. 

Next, we calculate the $\lambda_{\rm mpf}$ predicted by our models. We define,
%
\begin{figure}
	\centering
	\includegraphics[bb=93 362  510 710,width=8.0cm,keepaspectratio] {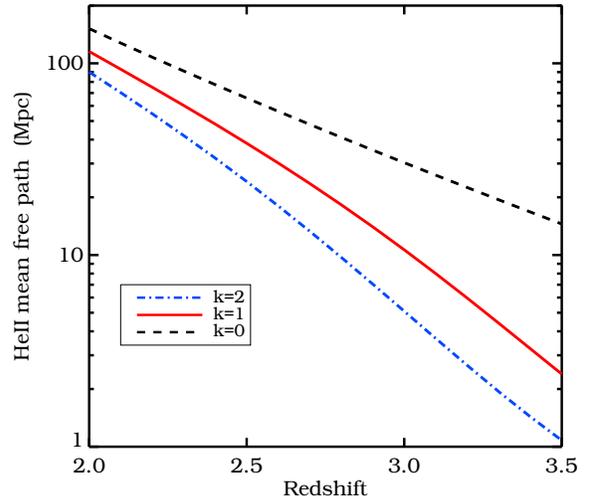}
	\caption{ \small Mean free path for He~{\sc ii} ionizing photons as a function of 
	$z$ for different $k$ values. Here, distance is in proper length scale.}\label{mfp}
\end{figure}
\begin{equation}
\lambda_{\rm mfp}=\bigg{|}\frac{dl}{dz}\bigg{|} \quad \frac{dz}{d\tau_{\rm  eff}} \,\,,
\end{equation}
%
where, from Eq.~(\ref{tau_eff}), we write,
%
\begin{equation}
\frac{d\tau_{\rm eff}}{dz}=\int_{0}^{\infty} {\rm dN_{\rm HI}} \, f(N_{\rm HI},\, z) \, \big[1-e^{-\tau(228\mathring{\rm A})}\big]\,\,\,.
\end{equation}
Here, $\rm \tau(228\AA)=\rm N_{\rm HI}(\sigma_{\rm HI, 228\mathring{\rm A}}+
\eta \sigma_{\rm HeII, 228\mathring{\rm A}})$. 
In Fig.~\ref{mfp}, $\lambda_{\rm mfp}$ of \HeIIa ionizing photons 
is plotted as a function of $z$ for different values of $f_{\rm esc}$.
These curves are well approximated by $\lambda_{\rm mpf}(z) = A_1 \times exp[-(z-2)/\Delta z]$ 
with best-fitting values of A$_1$=[150, 120, 90] Mpc and 
$\Delta z$=[0.60, 0.41, 0.37], respectively, for $k$=[0,1,2]. 
This clearly demonstrates
the rapid reduction in  $\lambda_{\rm mpf}$ with $f_{\rm esc}$. This is a 
consequence of the fact that when we increase $f_{\rm esc}$, 
a given $N_{\rm HeII}$ is produced by a cloud with lower 
$N_{\rm HI}$ (i.e. $\eta$ increases). 
Because $f(N_{\rm HI},z)$ is a powerlaw in $N_{\rm HI}$ with a negative
slope, $\lambda_{\rm mfp}$ reduces steeply. For our fiducial model
with $k\,$=$\,1$,  $\lambda_{\rm mfp}$ is 22 and 11 proper Mpc at $z\,$=$\,2.7$ and 3.0, respectively. These values are at least a factor 2 smaller than the 
corresponding values for $k\,$=$\,0$.

Recently, \citet{DF12} have computed the mean evolution of 
$\tau_{\alpha,{\rm HeII}}$, assuming diffuse emissivity of QSOs and 
allowing for fluctuating $\Gamma_{\rm HeII}$. This model also
produces a rapidly evolving  $\tau_{\alpha,{\rm HeII}}$ with $z$
without implicitly assuming He~{\sc ii} reionization around $z\sim3$,
when a minimum $\lambda_{\rm mfp}$ of 35 comoving Mpc is used.
This is already more than $\lambda_{\rm mfp}$ we obtain at $z=3$ for
our fiducial model. Therefore, we can conclude that $f_{\rm esc}$
will play an important
role in the calculations of the fluctuation. We also speculate that 
the redshift at which the fluctuations begin to dominate will
depend on $f_{\rm esc}$. Detailed investigations of high
ionization species, such as N~{\sc v} and O~{\sc vi} absorption in
QSO spectra at $z\ge2.5$, and proximity effect analysis of \HeIIb
Ly-$\alpha$ forest will provide more insights into the issues discussed here.
We plan to address these topics carefully in the near future.

\section{Conclusions}\label{sec.dis}

We have studied the effect of escape fraction of \HIa ionizing photons
from high-$z$ galaxies on the UVB, calculated by solving cosmological
radiative transfer equation.
We have demonstrated that \HIa ionizing photons from galaxies
play an important role in deciding the shape of \HeIIb ionizing 
part of UVB and that they affect the transmission of the UVB through the
IGM. Here, we summarize our main results.

(i) The He~{\sc ii} ionizing part of UVB depends greatly 
        on the value of $f_{\rm esc}$ for $z>2.0$.
	This is a consequence of the dependence of ratio 
	$\eta\,$=$\,N_{\rm HeII}/N_{\rm HI}$ on $f_{\rm esc}$.

(ii) Mean free path of He~{\sc ii} ionizing photons decreases 
        rapidly with increasing
        $f_{\rm esc}$, which suggests that it will play an important role in
        quantifying the fluctuations in He~{\sc ii} ionizing UVB.

(iii) We show that, for the range of $f_{\rm esc}$ allowed by the observations, 
        the rapid increase in He~{\sc ii} Ly$\alpha$ effective 
        optical depth (recently 
	observed at $z\,$$\sim\,$2.7) can be explained naturally. 
        In the literature such a trend is 
        attributed to the pre-overlap era of He~{\sc iii} bubbles, 
        assuming that the
        \HeIIa re-ionization completes around $z\,$=$\,2.7$. 
        Our study, while providing an alternate explanation, 
        does not rule out this possibility.
        However, we show that it will be possible to 
        place additional constraints
        on $f_{\rm esc}$ in the post-He {\sc ii} reionization era
        using well measured He~{\sc ii} Ly$\alpha$ effective optical depths.
        We also show that the observations at $z\le 2.7$ 
        are consistent with the fact that 
        $f_{\rm esc}$ is not much higher than 8\%.

\section*{acknowledgments} 
We wish to thank T. R. Choudhury, K. Subramanian  and the referee for useful  suggestions. 
VK thanks CSIR for providing support for this work.

\appendix
\def\apj{ApJ}%
\def\mnras{MNRAS}%
\def\aap{A\&A}%
\def\apjl{ApJ}
\def\aj{AJ}
\def\physrep{PhR}
\def\apjs{ApJS}
\def\pasa{PASA}
\def\pasj{PASJ}
\def\pasp{PASP}
\def\nat{Natur}

\bibliographystyle{mn2e}
\bibliography{vikrambib}

\end{document}